**Title:** Mean Field Theory Based on the Spike Time Response Curve for Synchronization Within and Between Two Alternating Populations of Neural Oscillators with Delays

**Authors:** Ananth Vedururu Srinivas and Carmen C Canavier*

**Affiliation:** Dept. of Cell Biology and Anatomy, Louisiana State University Health Sciences Center – New Orleans, New Orleans LA 70112

* Corresponding Author

Email: ccanav@lsuhsc.edu

**Abstract**

In order to study oscillators coupled by pulses emitted at discrete times, it is generally assumed that the effect of each pulse is an instantaneous phase shift, with a possible conduction delay between emission and receipt. However, we are interested in neural oscillators in which each pulse (action potential) activates a biexponential synapse, and at high frequencies, the duration of the synaptic conductance may be greater than the network period. Therefore, the effect of received pulses can summate over the course of several cycles. To overcome this limitation, we previously defined a mean field approach for a single synchronous population. Here, we extend it to two alternating synchronous populations by assuming a perturbation of a single oscillator from one of the synchronous populations which is represented by a single self-connected neural oscillator. The train of delayed biexponential synapses emitted by each oscillator is divided into a tonic and a phasic component. Because of the continuous rather than pulsatile nature of the input, we had to drop the concept of oscillator phase and derive self-consistent criteria for the existence and stability of an alternating firing pattern between two synchronous populations based solely on the perturbation of time intervals from their steady state values. Simulations showed that for high frequency oscillations as defined above, the mean field approach predicts the existence and stability of phase-locking between two synchronous populations better than the instantaneous phase shift approach. This approach may generalize to other forms of coupling amongst non-neural oscillators.

## Introduction

The study of pulse-coupled oscillators (Winfree, 1990) was popularized by Arthur Winfree. The basic idea is that the phase of each oscillator continuously updates at a constant frequency except for discrete jumps at the times at which a pulsatile input is received. Several different proofs of synchrony amongst identical oscillators coupled with no delay have been derived under different assumptions about the form of the pulsatile coupling and under different types of perturbations (Peskin, 1975; Mirollo and Strogatz, 1990; Goel and Ermentrout, 2002; Achuthan and Canavier, 2009). However, biologically realistic coupling is not instantaneous, and an input delivered late in the cycle may affect the next cycle. This realization led to an approach in which the resetting that occurs in the cycle after the perturbation was called second order resetting, which was successfully applied in some cases (Oprisan et al., 2004). However, under that formalism there was no way to account for effects of a pulse that lasted more than a single network period, therefore we developed a mean field approach (Canavier, 2025) to account for the effects of a train of perturbations on all subsequent cycles. This theoretical approach is more mechanistic and complete than our previous approach (Cui et al., 2009), which was simply to apply the train and look for phase-locked modes. We also account for conduction delays, which can change the stability criteria (Woodman and Canavier, 2011; Vedururu Srinivas and Canavier, 2024).

Here we present a mean field analysis of two populations of neural oscillators reciprocally coupled via biexponential synaptic conductances with conduction delays. Our previous work on the mean field of coupled oscillator networks only addressed a single population of neurons. Here we address the case of two distinct populations, consistent with experimental observations that in many cases, two populations of neurons, one excitatory and one inhibitory, are required to generate gamma oscillations in the brain (Bartos et al., 2007). The choice of circuit in which to test our theoretical results numerically was motivated by our previous work in the superficial layers of the medial entorhinal cortex (Via et al., 2022) . We were also motivated by observations that a population of inhibitory interneurons that fired asynchronously in the absence of connectivity with a population of excitatory neurons (Pastoll et al., 2013) synchronized at gamma frequency in the presence of reciprocal connectivity with the excitatory population. In that particular example, the inhibitory cells have recurrent connectivity, but the excitatory cells (specifically the stellate cells in the medial entorhinal cortex) do not, as in the simulations we present in this study.

**Methods**

*Mean Field Response Curves*

Fig. 1A1 illustrates a periodically spiking neuron (top trace) receiving a periodic train of biexponential synaptic conductance inputs at the times indicated by the arrows (bottom trace). We define the stimulus interval ts as the interval between a spike in a neuron and the receipt of the next input. In a one-to one phase locked mode for two oscillators, each oscillator receives an input at a fixed interval ts* after it emits a spike, where ts* denotes the steady state value. The idea behind the mean field is that is the periodic input can be divided into a tonic component (horizontal dashed line in Fig. 1A1) and a phasic component. For a train of biexponential synaptic conductances extending infinitely into the past, the tonic component is taken as the minimum value of the sum of an infinite train of biexponential synaptic conductances given by $g_{tonic} = f\left(\frac{1}{1-e^{-P_F/\tau_d}} - \frac{1}{1-e^{-P_F/\tau_R}}\right)$ as explained in (Canavier, 2025), where f is a scale factor that normalizes the peak to 1 (Dayan and Abbott, 2005), $\tau_d$ is the decay time constant and $\tau_R$ approximates the rise time constant of the biexponential synapses if $\tau_R >> \tau_d$. We now define a mean field response under a mean field assumption, analogous to a phase response curve under the assumption of pulsatile coupling, that is, the spike time response curve (Acker et al., 2003; Canavier and Achuthan, 2010). We assume that the oscillator has a new free running period ($P_E$) due to the tonic component, (horizontal lines in the second trace from the top in Fig. 1A2), then we apply a single cycle of the waveform of the phasic input shown between the vertical dashed lines in the same panel. The arrow indicates that the input is received at stimulus interval ts*. The phasic part of the waveform is given by $g_{phasic}(t) = f\left(\frac{e^{-t/\tau_d}}{1-e^{-P_F/\tau_d}} - \frac{e^{-t/\tau_R}}{1-e^{-P_F/\tau_R}} - \frac{1}{1-e^{-P_F/\tau_d}} + \frac{1}{1-e^{-P_F/\tau_R}}\right)$ (Canavier, 2025). For the mean field response, prior to the trough at t=*ts* the phasic part of the synaptic conductance waveform results from an input prior to the current cycle that occurred at time ts- $P_F$ ms prior to the most recent spike that occurs before the input at the arrow and is given by $g_{phasic}(t - ts + P_F)\, g_{phasic}(t-ts+P_F)$. At the time an input is received, the value for the phasic part of the input is zero, resulting in a trough. After the trough for t = [ts, $P_F$), the phasic synaptic conductance waveform is due to the input applied at the trough and is given by $g_{phasic}(t - ts)\, g_{phasic}(t-ts)$. The response interval tr is the time interval after an input is received until the neuron fires next, thus the tr measured for ts* gives us the perturbed period $P_1$ as in the top trace.

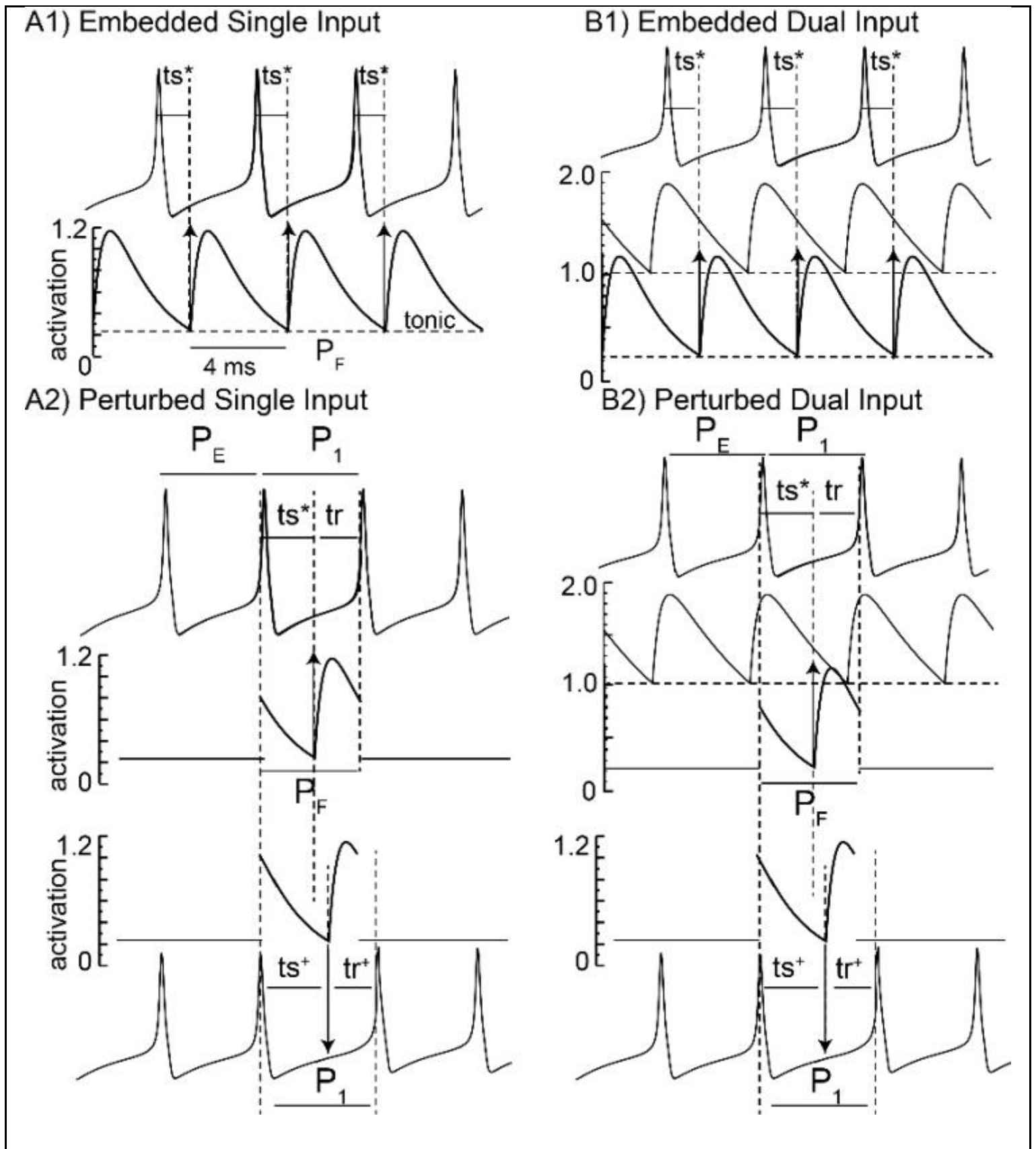

Figure 1. Mean Field Response Schematic. A1. Oscillator embedded in a periodic train of biexponential inputs from another neuronal oscillator. A2. Measuring the mean field response to a single phasic input. B1. Oscillator embedded in two periodic trains of biexponential inputs. A2. Measuring the mean field response to a single phasic input whilst keeping the other constant. See text for definitions.

The mean field response M(ts) is the difference between the length of the perturbed cycle $P_1$ and that of the cycle ($P_E$) of the oscillator embedded in the tonic conductance: M(ts) = $P_E$-$P_1$, where $P_1$ is the perturbed cycle period. We then assume that the embedded oscillator shifts its firing within the network slightly with respect to its usual firing pattern to measure how the mean field response changes with respect to shift in the firing pattern. This shift will be critical in determining stability, as we demonstrate in the results. Caveat: This approach neglects the effect of the previous phasic cycles on the state of the model neuron at the time a spike is emitted prior to application of the phasic waveform; this approximation simplifies the procedure but may introduce some inaccuracy into the method.

For the case of three oscillators, the perturbed oscillator is embedded in two tonic synaptic conductances and receives two phasic inputs per cycle at known time intervals as in Fig. 1B1. Synaptic conductances with the same reversal potential can be added but those with a different reversal potential cannot be added, unlike synaptic currents, because of the difference in driving force. However, we can still measure the mean field response by applying one train of synaptic inputs along with the two tonic inputs and one phasic input to simulate the network embedding, then apply and shift the other single phasic cycle as before (Fig. 1B2). The caveat is that we no longer have an autonomous limit cycle, but this is mitigated because we do not need the concept of phase along a limit cycle.

*Existence of 1:1 locking between two clusters*

Since we do not expect the two populations of oscillators to be synchronous, we assume an alternating firing pattern (Fig. 2A) between the clusters with transmission delays as in (Woodman and Canavier, 2011; Vedururu Srinivas and Canavier, 2024). . As in figure 1, the stimulus interval (ts) is defined as the interval between the time a neuron emits a spike and when it next receives an input (the trough of the phasic component in the mean field case). However, in Fig. 1 these quantities were defined in the open loop condition for measuring the mean field response such that connectivity is unidirectional. In Fig. 2A, these terms indicate the intervals in a 1:1 steady locked mode. The response interval (tr) is defined as the interval between the receipt of an input and the emission of the next spike; thus, the sum of these two intervals is the network period. Here we assume that the reciprocal delays between populations are symmetric and denote the external conduction $\delta_{ext}$ to differentiate them from the conduction delays within a cluster δ. In order to predict the activity of two synchronized clusters coupled with conduction delays, we use the algebraic constraint (Fig 2A) that in an alternating firing mode (Fig. 2C), the stimulus interval in one cluster is equal to the response interval in the other plus twice the external delay, and vice versa.

In order to determine the existence of a steady one-to-one alternating phase-locked mode, we first need to determine the predicted network period for each population oscillator by assuming a range of forcing periods by the other population oscillator for each stimulus interval. Note that the exact length of the perturbed period $P_1$ in Figures 1A2 and B2 does not necessarily match the duration of a single cycle of the assumed forcing period $P_F$, but these must match in a steady phase-locked mode. For each stimulus interval for each oscillator, each forcing period will result in a different embedded period at the tonic conductance and a different $P_1$ generated by the mean field response to the phasic conductance waveform as described in Fig. 1. The perturbed interval is also the predicted network period $P_N$. The existence of one-to-one lockings is given by the intersection of the predicted network period as a function of the forcing period with the diagonal on which the two quantities are equal (Fig. 2B). Then the phasic relationships of any phase-locked mode that exists will be determined by the intersection of the curves based on the algebraic relationships between the stimulus and response intervals determined using the mean field approach. We can find the response interval for the predicted1:1 locked mode at each external delay by subtracting two delay intervals from the network period ($tr^*=P_N-2\delta_{ext}$) and then plot the curves as in Fig. 2C).

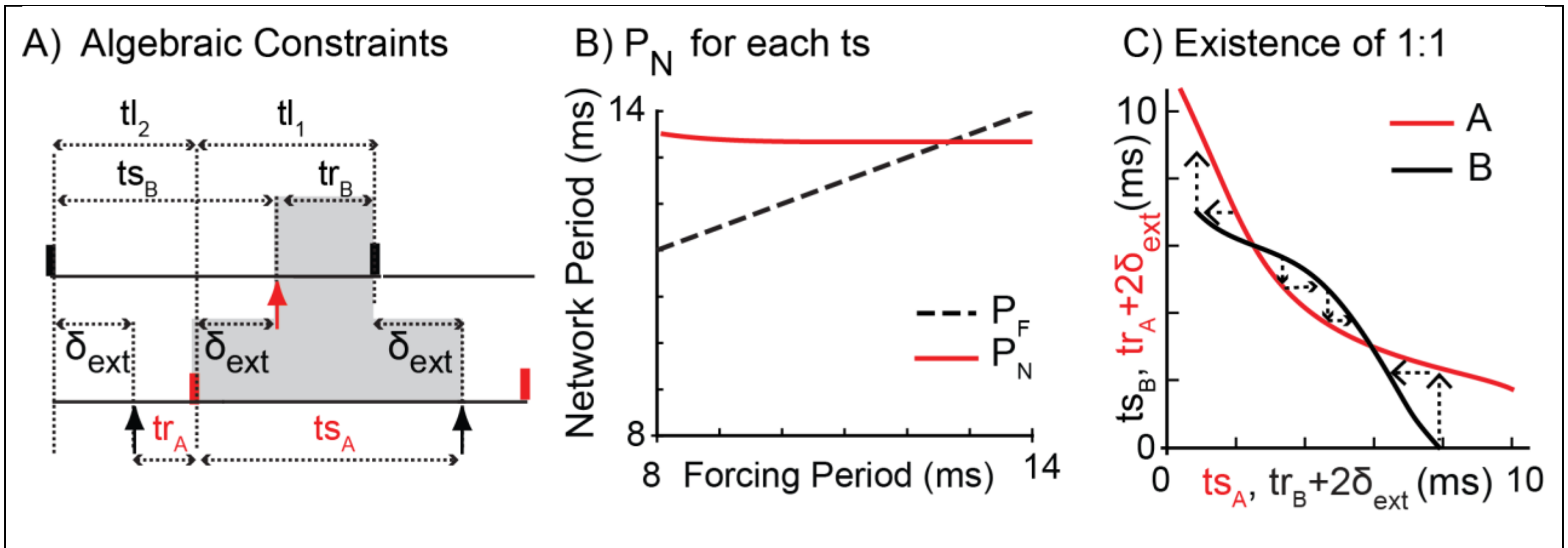

Fig 2. Alternating Firing Pattern with Delays. A) Algebraic constraints. The observable time lags between firing of one oscillator and firing of the next are denoted $tl_1$ and $tl_2$. B. Determination of the forcing period of one oscillator that matches the predicted network period. C. Determination of the phasic relationships for two alternating oscillators using the response intervals determined by $P_N$-ts in B plugged into the constraints from A. Since the stimulus interval in one neuron on any given cycle is equal to the response interval plus two external delays, the plot in this panel lends itself to a discrete map of the intervals (arrows).

*Neuronal models for theoretical analysis and Simulations:*

As an example to test our methods, we used simulations of a recurrently connected inhibitory (I) cell population reciprocally connected to an excitatory (E) cell population that does not have recurrent connectivity (Williams et al., 2026), so we represented the E population by a single cell.

We used conductance-based Hodgkin-Huxley type model neurons representing the inhibitory network cluster and the excitatory cluster.

For the network simulations, we modified the code repository used in (Canavier, 2025) to include an excitatory RTM neuron connected to the network of 100 homogenous inhibitory neurons with each neuron receiving exactly 36 synapses from other inhibitory neurons to eliminate jitter caused by random connectivity. Since this is equivalent to a network of 37 homogeneous neurons (a neuron is not connected to itself), for the theoretical work we assume the number of neurons N is 37. The internal delay δ for the I-I connections was always 0.3 ms. Compared to the inhibitory neuron model used in our previous studies (Via et al., 2022; Canavier, 2025), we doubled the capacitance for the inhibitory interneurons to increase the intrinsic free-running period. The E-I and I-E synaptic delays were made equal for convenience. We used both hyperpolarizing ($E_{REV}$ = -75 mV) and shunting ($E_{REV}$ = -55 mV) synapses for the I network. For the RTM neuron, we used an external current drive of 110 pA when the I-I synapses were shunting and 60 pA when the I-I synapses were hyperpolarizing.

For the I-E synapses, we used $E_{REV}$ = -75 mV, with a rise time constant of 0.3 ms and decay constant of 5 ms. The total inhibitory synaptic conductance received by an RTM neuron is 2.75 nS. For the E-I synapses, we used $E_{REV}$ = 0 mV, with a rise time constant of 0.3 ms and decay constant of 2 ms. The maximum excitatory synaptic conductance received by an inhibitory interneuron is 9 nS.

For our theoretical predictions, we used a self-connected neuron to represent the inhibitory network cluster minus 1 neuron that was represented separately, and a single RTM model to represent a synchronous excitatory cluster that is not self-connected, both of which are described in detail in our previous work (Vedururu Srinivas and Canavier, 2024).

We used python notebooks and the python-based BRIAN2 simulator (Stimberg et al., 2019). For running large scale parameter sweeps, we used HPC clusters to parallelize the simulations. The codes for all the simulations can be found in the repository git@github.com:AnanthVS23/Meanfield_EI.git.

**Results**

*Development of new theory*

In pulsatile coupling theory, the phase is expected to increase linearly with elapsed time during the stimulus interval (ts) between the time when a neuron emits a spike and when it receives the next input. In our protocol for measuring the response of an oscillator to a perturbation, the part of the phasic input due to the previous spike of the other oscillator is active during the stimulus interval, the phase is not defined during the stimulus interval. Therefore, we drop the notion of phase and simply measure difference in cycle length due to perturbation at any given stimulus interval. The difference in cycle length is, of course, contained entirely within the response interval.

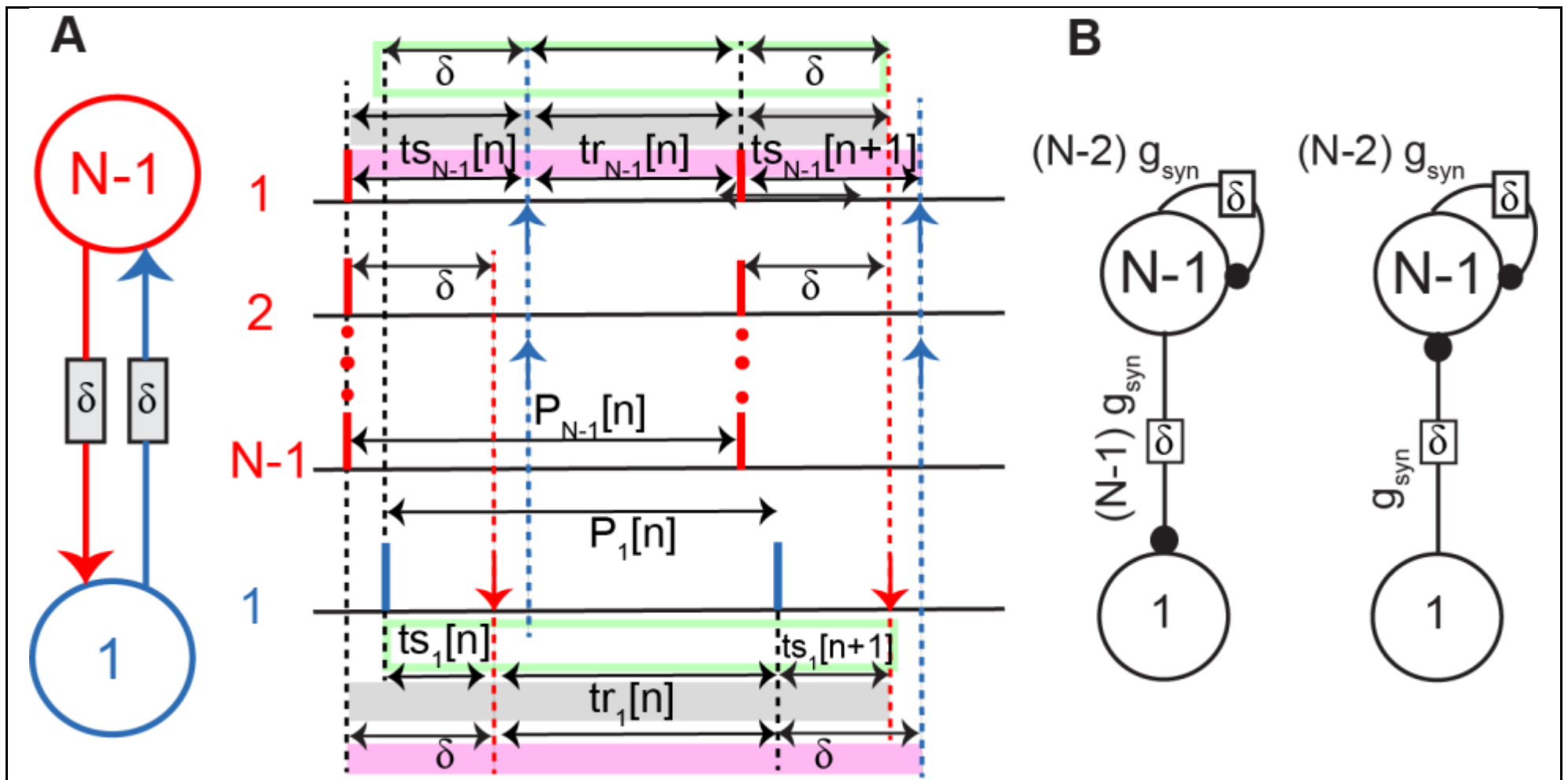

**Fig. 3. Stability of Synchrony in a Network of N identical neurons each receiving N-1 identical inputs.** A. Closed loop circuit (left). Schematic firing pattern in which a single neuron is perturbed from synchrony (right). These intervals are used in the proof of stability(Woodman and Canavier, 2011; Vedururu Srinivas and Canavier, 2024). B. Open loop IRCs schematic for reduction of the network to two oscillators. The larger group is assumed to receive delayed synchronous feedback from itself.

The rederivation of the proofs for the case of two alternating oscillators and for global synchrony is trivial, one merely has to omit the steps in which 1) the stimulus interval is normalized by the intrinsic period to obtain the phase and 2) the difference in cycle length is normalized by the intrinsic period to get the phase response curve. Instead of phase, all quantities in the derivations have units of time. The population is perturbed into two oscillators: one oscillator is considered to be a cluster of J neurons with J-1 self-connections and the other cluster of N neurons with N-J-1 self-connections. For the simplest case J = 1 (with no self-connections), we obtain the stability criterion:

$$|1 - M_1'(\delta) - M_{N-1}'(\delta)| < 1 \qquad (1)$$

where M denotes the interval response curve under mean field assumptions, the apostrophe denotes the derivative with respect to the stimulus interval ts, and δ represents the conduction delay for the connections within the cluster. More generally, we get

$$|1 - M_J'(\delta) - M_{N-J-1}'(\delta)| < 1$$

The algebraic constraints and the discrete maps for the result previously derived for pulsatile coupling with delays (Vedururu Srinivas and Canavier, 2024) are described in Appendix A for the stability of all-to-all synchrony case.

*Stability of phase-locking between two alternating clusters*

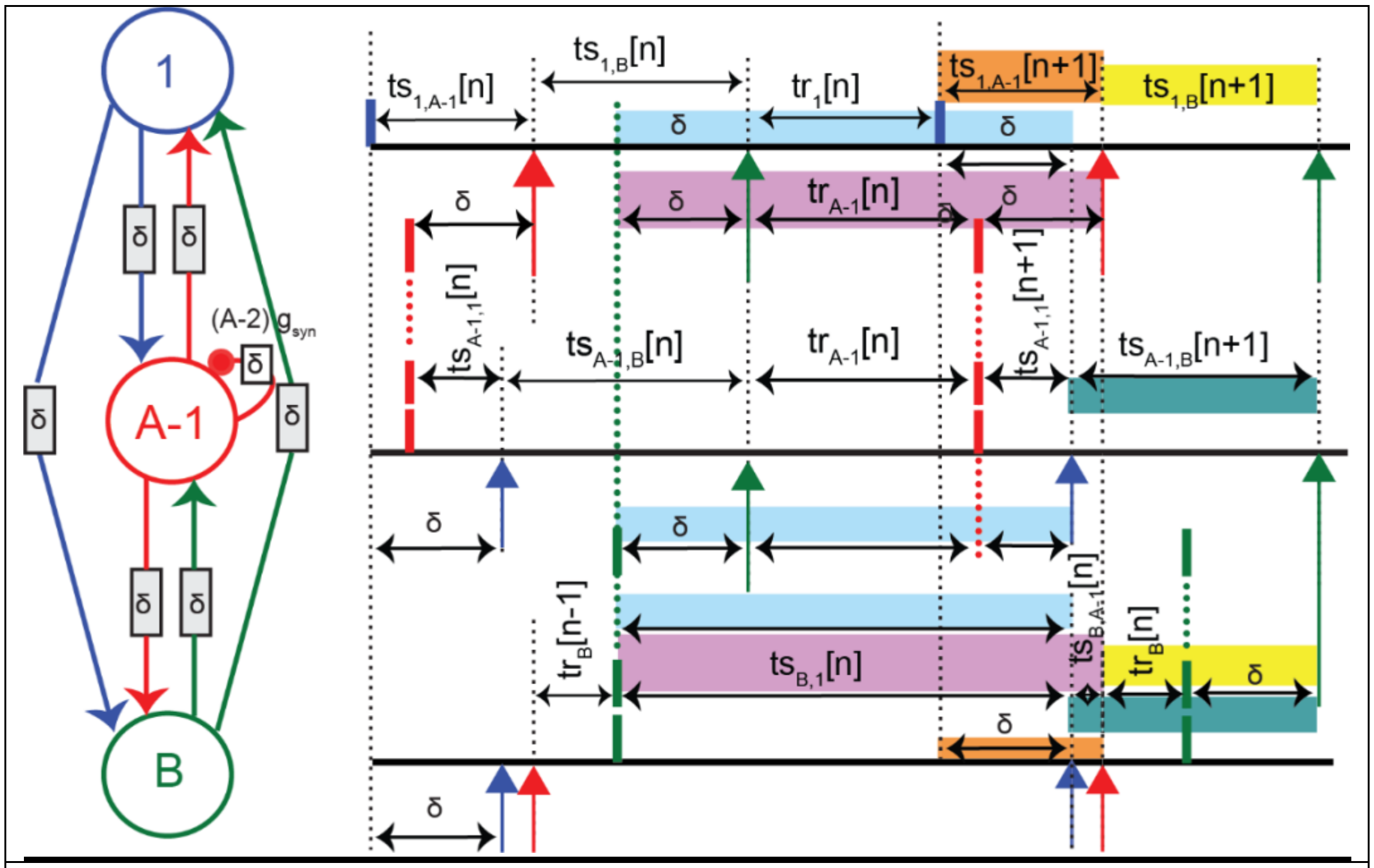

**Fig. 4. Stability of Synchrony in Two clusters with alternating firing pattern.** Firing pattern and intervals when a single neuron is perturbed from the self-connected cluster in a network with two alternating clusters. The cluster B is not self-connected. Intervals depicted using the same color are equal.

For the case of two alternating populations that are synchronized within each population, the proof is more challenging because in the perturbed condition, there are three oscillators and each oscillator receives two inputs per cycle. There are three oscillators because one of the populations is again perturbed into two oscillators: a single oscillator, and a self-connected oscillator that represents the remainder of the population. Here, we measure the effect of shifting one cycle of phasic input not only in the presence of both tonic inputs but also in the presence of the other continuous train of phasic inputs. In the case of instantaneous pulsatile coupling, one can merely assume that the phase of the oscillator at the time the second input is received has been adjusted by the phase resetting caused by the first input. Here we refer to two stimulus intervals (ts1 and ts2): The first stimulus interval begins when a spike is emitted; The second stimulus interval begins when the first input is received and ends when the second stimulus is received. The response interval begins when the second stimulus is received and ends when a spike is emitted. In order to construct a map of the intervals, we make the assumption that the change in cycle length due to the first input is entirely contained in the second stimulus interval and adjusts the effective time at which the second input is received. Since the effect of the first stimulus can potentially bleed over into the response interval, this assumption introduces a possible source of error, but our simulation results (below) validate this assumption.

For the existence criterion described in Fig. 2, only two mean field responses need to be measured: $M_{A,B}(ts_A)$ and $M_{B,A}(ts_B)$, where the first subscript of the mean field response indicates the oscillator receiving the input and the second the oscillator providing the input. However, for the stability criteria, there are six mean field responses with a third subscript. Only the slopes around the steady state value are required so we will only indicate the slopes. We assume the internal delays in cluster A are constant, so we do not need a subscript for δ. For the dual input protocol, we require $M'_{A-1,1,B}(\delta)$, $M'_{1,A-11,B}(\delta)$ , $M'_{B,A-1,1}(ts_B^*)$, $M'_{B,1,A-1}(ts_B^*)$, $M'_{A-1,B,1}(ts_A^*)$ and $M'_{A-1,B,1}(ts_A^*)$, where the third input indicates embedding in the continuous biexponential train of the oscillator indicated by the third input. However, under the mean input assumptions, $M'_{A,B}(ts_A^*)$ is exactly equal to $M'_{A-1,B,1}(ts_A^*)$ and $M'_{1,B,A-1}(ts_A^*)$ because $tr_A$ is the same for all three cases due to the dual embedding, thus we can substitute $M'_{A,B}(ts_A^*)$ for $M'_{A-1,B,1}(ts_A^*)$ and $M'_{A-1,B,1}(ts_A^*)$. In Appendix B, we rederive the map for the alternating cluster case as our previous work (Vedururu Srinivas and Canavier, 2024), but using the mean field approach. We make the further simplifying assumption that only cluster A has recurrent connectivity (otherwise the eigenvalues below must also be evaluated with the A and B indices reversed). Then, the two eigenvalues are given by

$$\left(1 - M'_{A,B}(ts_A^*)\right)\left(1 - M'_{1,A-1,B}(\delta) - M_{A-1,1,B}(\delta)\right) = \lambda 1 \qquad (2)$$

$$(1 - M'_{A,B}(ts_A^*))(1 - M'_{B,A-1,1}(ts_B^*))((1 - M_{B,1,A-1}(ts_B^*)) = \lambda 2 \qquad (3)$$

Note that the derivation assumes short delays, meaning the external delay must be shorter than the shortest time lag in Fig. 2A so that $\delta_{ext} < \min(tl_1, tl_2)$. Otherwise, the target cluster fires after spike emission in the presynaptic cluster but before the input reaches the target cluster. For longer delays different stability and existence criteria apply, and in general stability decreases as delays increase (Woodman and Canavier, 2011).

*Simulations to test the theory: fast gamma oscillation range with hyperpolarizing synapses*

We calculated mean field response curves on the RTM and PV-IN models described in the Methods for two cases involving I-I synapses: hyperpolarizing synapses with $E_{REV}$ = -75 mV and shunting synapses with $E_{REV}$ = -55 mV. In both cases, the I cells were connected to each other with an internal conduction delay δ of 0.3 ms. The rationale was to illustrate fast oscillations in which inputs may summate, such as in the fast gamma range (65-140 Hz) (Colgin et al., 2009) as in Fig. 5 or in the range of ripple oscillations (140-200 Hz) in the hippocampus (Csicsvari et al., 1999) as in Fig 6 and 7. In these examples, cluster A is the I cells and cluster B is the E cell.

When the I-I synapses are strongly hyperpolarizing, the I cluster can synchronize by itself at most delays less than the intrinsic period of the I cluster ( Vedururu Srinivas and Canavier, 2024, Canavier, 2025). When we connect the E cell to the I cluster, the network frequency increases and a one-to-one (1:1) locking is observed based on the self-consistency criterion and the existence criterion (as shown in Fig 2B and 2C). Based on the existence of 1:1 locking for multiple delays, we observed that the smallest time lag between the firing of the I cluster and the E cell is about 2.18 ms. Therefore, we varied the E-I delay from 0.1 to 2.15 ms to check the robustness of the mean field predictions, while honoring the $\delta_{ext} < \min(tl_1, tl_2)$ and keeping all other parameters the same. Fig. 5 shows the results obtained through theoretical predictions along with the observations from the network simulations. For each delay $\delta_{ext}$ in the interval [0.1, 1.60], the tr-ts curves intersected at two points resulting in two possible 1:1 lockings. If there are only two fixed points of the map (the intersections in Fig. 2C) it is obvious that only one of them can be stable (Thounaojam et al., 2014).

In Fig. 5, tl1 indicates the interval between the time that I population fires and the time that the E cell fires next, whereas tl2 indicates the interval between the time that E cell fires and the time that the I population fires next. From Fig. 2A, we see that at steady state, the network period is given by the sum of lags tl1+tl2, or equivalently by the sum the stimulus and response intervals of each oscillator: Network Period = tl1+tl2 = $ts^*_A$ + $tr^*_A$ = $ts^*_B$ + $tr^*_B$. We use the time lags for plotting because they give the firing pattern. Thus, if tl1 and tl2 are equal, the pattern is antiphase firing. If tl1>tl2 then the E cell fires first, and if tl1<tl2 then the I population fires first. In Fig. 5A2, at all delays in which a stable mode is observed, the E-cell fires first (open triangles). The key observation is that $ts_A = tl_1 + \delta_{ext}$ (see Fig. 2A). In Fig 5A2, most values of tl1 are ~5.5, for as $ts_A$ of ~between 5.5 and 7.0 ms, which falls in the region of positive slope for $M_{A,B}(ts_A)$ (the response of the I cell cluster to an input from the E cell) in Fig 5C. Since this slope is less than one, it is highly stabilizing. As the delays are increased, the model continues to predict 1:1 locking (to the right of the vertical dashed line) at a few values at which locking was not observed; either the assumptions break down in this small region, or the basin of attraction of the stable mode was too small for us to detect it. At delays greater than 1.65 ms, there are no intersections, thus no 1:1 locked modes are predicted to exist, nor were any observed. The predicted frequencies for the stable modes are indistinguishable from the observed frequencies, and the predicted times lags have only a very small error. In Fig. 5B, in all the unstable modes that are predicted to exist, the I cells fire first with tl1 of ~2.0 implying a $ts_A$ between 2 and 3.5 ms, which falls in the region of negative slope for $M_{A,B}(ts_A)$ in Fig. 5D. Moreover, $ts_B = tl_2 + \delta_{ext}$ (see Fig. 2A), so $ts^*_B$ is around 6.0 ms from Fig. 5B, and the slope of the corresponding mean field response in Fig. 5D2 is also negative and therefore destabilizing._

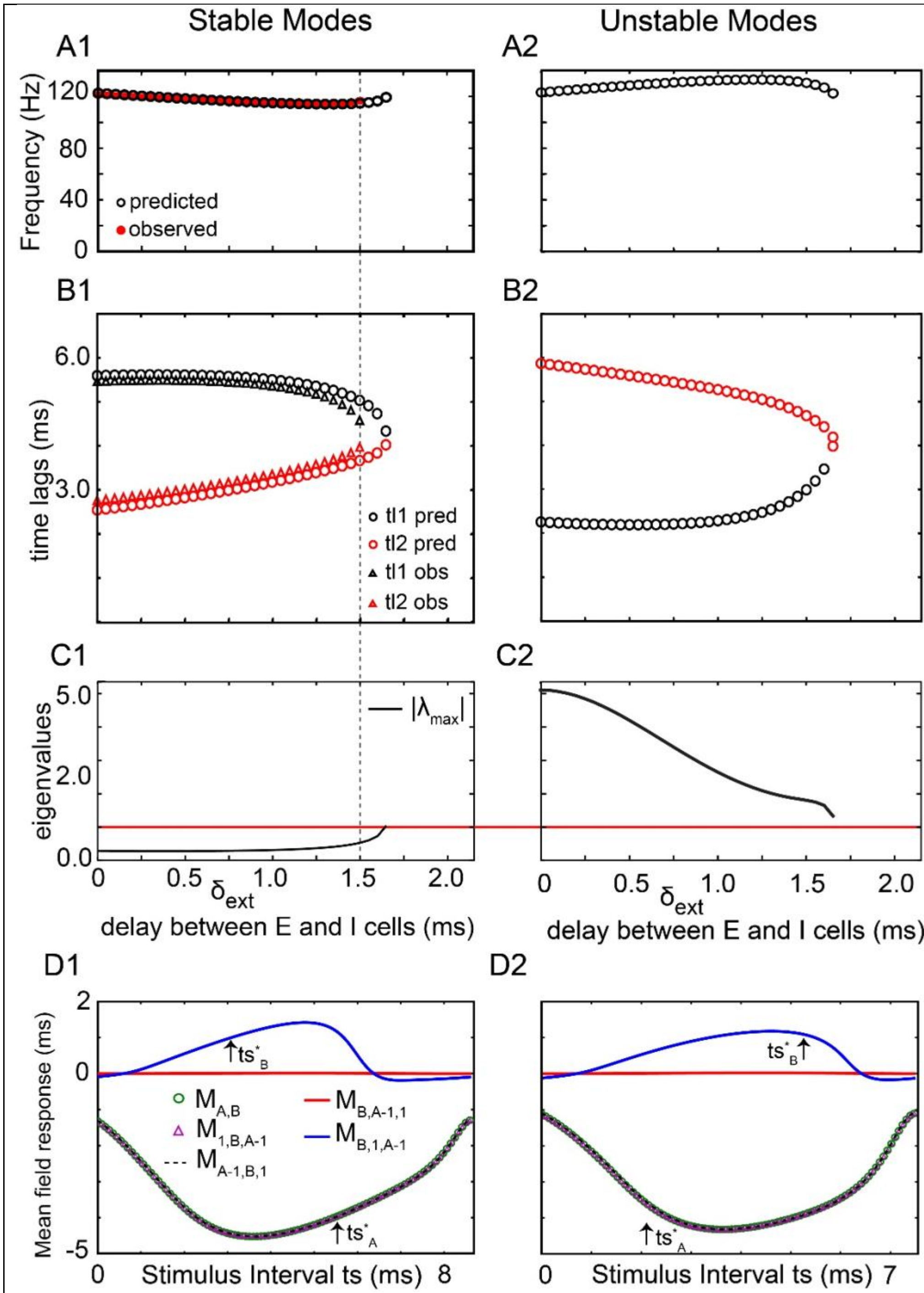

Fig. 5 Mean Field Method Correctly Predicts time lags and stability in Fast Gamma range 110-120 Hz. Left: Predicted stable modes, Right: Predicted unstable modes. A. Network frequency. B. Time lags, tl1 is interval between I cell firing and E cell firing, tl2 is between E cell and I cell firing. C. Maximum absolute value eigenvalue. Red line at one indicates stability boundary. Predicted modes to the right of the vertical dashed line were not observed. D. Mean field response curves with the point at which the slope should be taken indicated by the arrows. Note the identity of the three responses to the inputs from the E cell (middle subscript B). The response of the E cell to the single I neuron (red) is negligible.

*Simulations to test the theory: ripple oscillation range with shunting synapses*

When the I-I synapses are shunting and not connected to E cells, the I cells cannot synchronize themselves (Vedururu Srinivas and Canavier, 2024; Canavier, 2025) in the range of physiologically measured delays (Fernandez et al., 2022; Via et al., 2022). As a result, even though all-to-all synchrony of the I cells exists (Canavier, 2025), it is not stable. To test this, we changed the reversal potential of the I-I synapses to shunting ($E_{REV}$ = -55 mV) and increased the drive to the E cells to 110pA for the example considered in Fig. 5. We varied the delay in the range 0.1 – 1.4 ms to honor the condition $\delta_{ext}< \min(tI_1,tI_2)$. Based on the criterion in Eqn. 1, the eigenvalue for the I network without E cells is 1.511 which is greater than 1 (see Fig. 5C). When connected to the E cells, the two clusters fire in an alternating 1:1 locking pattern and the network produces high-frequency oscillations (~190 Hz) (see Fig. 6A). IN the presence of E cells the maximum eigenvalue for all the delays remains well below 1 ensuring stability of the 1:1 locking. We discuss this in detail for a specific example with $\delta_{ext}$ = 0.1 ms in the following section.

Since the pulsatile approach has also been used to predict time lags and stability of synchrony, we contrast those predictions with results obtained using the mean field approach. Fig. 6B shows that the mean field predictions of time lags are much closer to the observed lags than those obtained using the pulsatile approach, validating the mean field approach for fast input frequences. Thus, we only compare the predicted network frequencies with the frequencies observed in the simulations in Fig. 6A. Fig 6C shows the maximum eigenvalues for the isolated I cell cluster (dot-dashed line) and the maximum eigenvalue for the coupled system (solid black curve). The predictions are only shown for $\delta_{ext}\leq 1.4$ ms because the time lags have that value, and the assumed firing pattern in Fig. 2A no longer holds if a neuron spikes after the partner spike that emitted an input but before the input is received.

In order to explain how reciprocal coupling to the E cell enforces synchrony in the I cell population, we consider a specific example from Fig. 6 for which $\delta_{ext}$ = 0.1 ms. Fig. 7A shows that when the I cells are connected to each other using shunting synapses but not connected to the E cells, they fire asynchronously when perturbed from perfect synchrony because synchrony within the cluster is unstable. The I-cell cluster cannot synchronize by itself because $\left(1 - M'_{1,A-1,B}(\delta) - M_{A-1,1,B}(\delta)\right) = 1.511 > 1$ (see Fig. 6C). When connected to the E cells, the two clusters fire in an alternating 1:1 locking pattern (see Fig. 7B) and the network produces high-frequency oscillations (~190 Hz). In order to understand why synchrony results, we need to examine the two intersections indicated by the arrows in the tr-ts curves in Fig. 7C. One of the intersections has a positive slope $M'_{A,B}(ts_A^*)$ of the mean field response in Fig 7D, which gives the response of the I cluster due to the input from the E cell at a fixed forcing period calculated from Fig. 7C for $\delta_{ext}$ = 0.1 ms. Eigenvalue $\lambda 1$ from equation 2 above consists of the term $\left(1 - M'_{A,B}(ts_A^*)\right)$ that multiplies the eigenvalue for within cluster synchrony $\left(1 - M'_{1,A-1,B}(\delta) - M_{A-1,1,B}(\delta)\right)$. Note that when we measure the mean field responses within the cluster in the

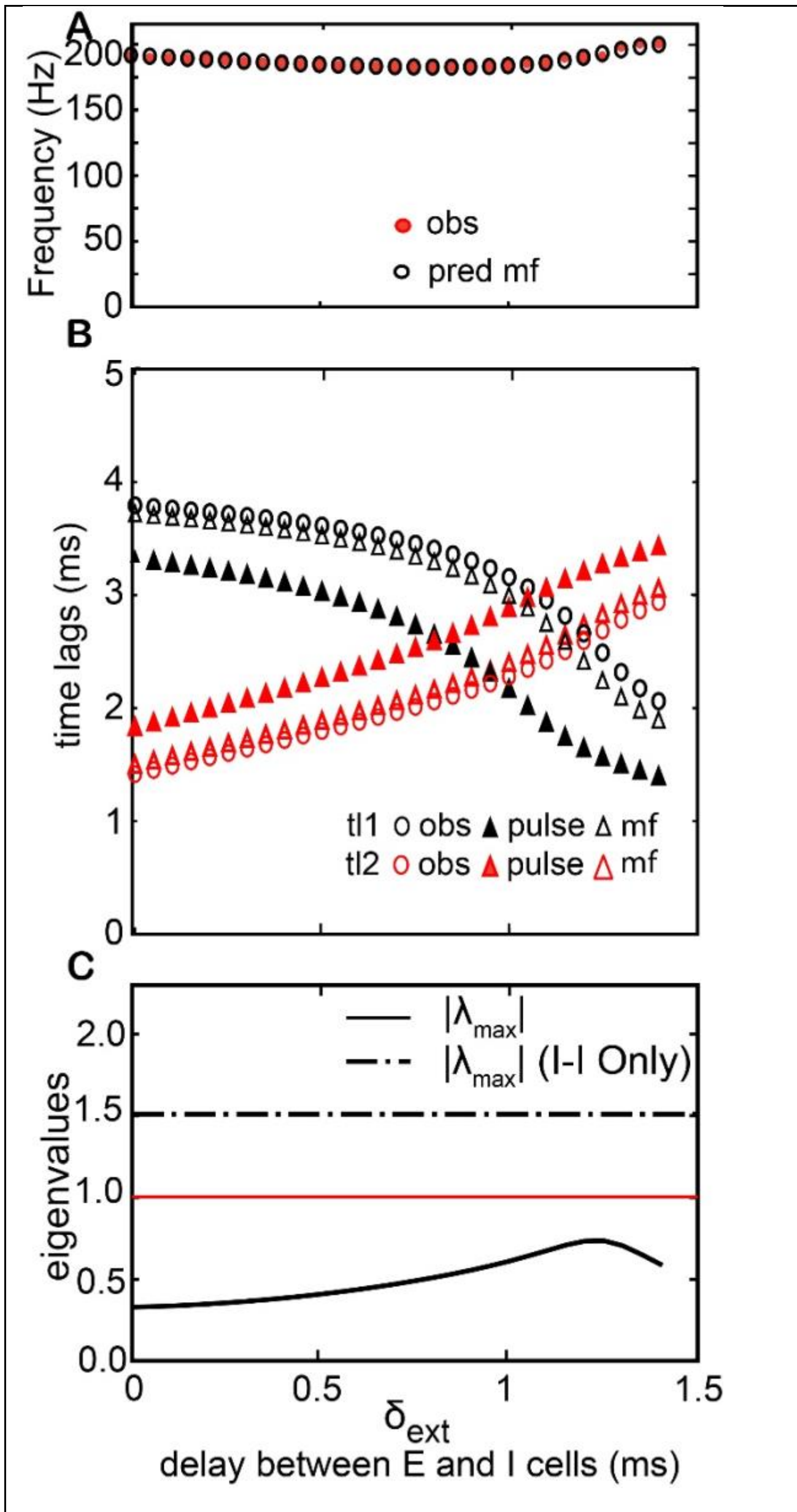

Fig. 6 Mean Field Method predicts Phasic Relationships and Stability better than the pulsatile approach. A. Network Frequencies. B. Times lags as defined for Fig. 5. C. Once again the red line at one indicates the stability boundary. Reciprocal coupling to the E cell stabilizes synchrony in the I cell cluster.

presence of E cells (using both phasic and tonic inputs), the value of the within cluster term is 1.09 (1.06) depending on the ts for input from E cell, which is still more than 1. This term due to the coupling with cluster B reduces the absolute value of the

eigenvalue to 0.5, which is less than one, thereby stabilizing the cluster that could not synchronize

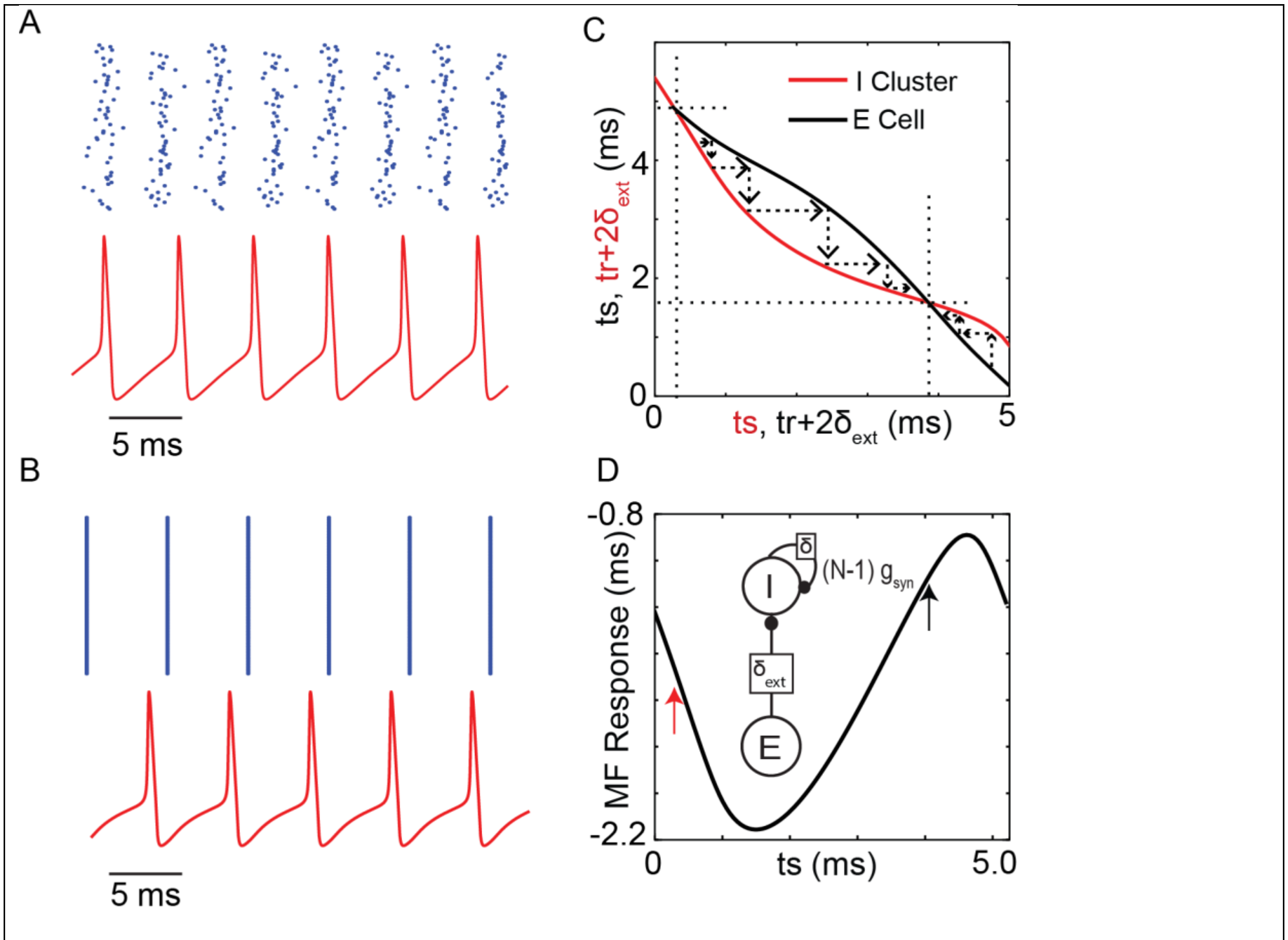

Fig 7. Reciprocal Coupling to E cell synchronizes I cell cluster: A) E cell and I cell network are not connected. The I cell cluster does not synchronize by itself. B) Connecting with E cell synchronizes the I cells in a 1:1 locking. The synaptic E-I delay is 0.1 ms. C) The tr-ts curves intersect at two points D) Mean field response of the I cluster for the input from the E cell. The locking point closer to 0 is unstable (red arrow) as the slope of the IRC is negative. The second intersection point is stable due to the positive slope (black arrow)

in isolation. For the unstable mode, the value of $M'_{A,B}(ts_A^*)$ is negative resulting in a higher $\lambda 1$ which is 1.72.

## Discussion

In this work, we used mean field approach to study the stability of all-to-all synchrony within a cluster and alternate firing pattern between clusters of neuronal oscillators. In our previous work (Canavier, 2025), we introduced the mean field approach and studied within cluster synchrony. Here extended the approach to include two neuronal populations and mathematically analyzed their stability by deriving the previously studies criterion using the new approach. Our previous work (Canavier, 2025) identified two extreme cases in which the mean field assumptions break down with overly strong inhibition.1) In the case of shunting inhibition, the inhibition cannot summate so strongly after two (or more) spikes that it essentially clamps the neuron at the shunting reversal potential as it decays during interval following the last input, resulting in alternating short and long intervals. 2) The hyperpolarizing inhibition cannot occur so late in the cycle that it fails to prevent a spike and then summates with the inhibition triggered by the next spike to strongly delay every other interval (Canavier, 2025).

For overly strong excitation, the I cells fire more than once for every spike from the E cell, leading to a 1:n firing pattern, where n is an integer. The methodology we use is not suitable for studying 1: n locking modes.

A previous mean field analysis of oscillations in neural populations assumed that neurons were subthreshold and that all spiking was fluctuation driven by noise (Brunel and Hansel, 2006). The point at which the stationary asynchronous stationary state gave way to a population oscillation was determined using a self-consistency criterion between the output of the system, which is the population firing rate, and the input, which is the train of biexponential synaptic waveforms triggered by the population firing rate. If there is an oscillation in the firing rate, this rate is transformed into an oscillation in current by the linear operators that characterize a biexponential synapse. This oscillatory current then drives into an oscillation in rate by the individual neurons receiving the oscillatory synaptic input. For an oscillation at a given frequency to exist, the oscillation in rate must have the same phase and amplitude before and after it is transformed into a current and back. The main point of that line of analysis was to explain how oscillations arose solely from synaptic interactions within a single population of neurons in the noisy, fluctuation driven regime, or between two populations in the same regime (Brunel and Wang, 2003). In contrast, we consider oscillators in the mean driven regime (Schreiber et al., 2009) in which they fire periodically.

A recent mean field formulation (Dumont et al., 2022) used a single refractory density equation to characterize a population of individually stochastic neurons represented by the amount of time passed by since the most recent action potential occurred ( a renewal process). Oscillations arise when the asynchronous solution loses stability. They then derived a macroscopic infinitesimal phase response curve for each neural population as a whole, and used weak coupling theory(Ermentrout and Terman, 2010) to predict phase-locking between two populations of neurons relies on weak coupling theory. In this formulation, as in the mean field approach described above and in contrast to the present study, individual neurons are not oscillators and oscillations arise only at the level of populations.

Although the theory of stochastic population rhythms has been very influential, there have always been some reservations that the firing of individual neurons might actually in some cases be oscillatory (Bartos et al., 2007). A recent paper (Huang et al., 2024) used optogenetics to record the spiking activity of PV+ inhibitory interneurons *in vivo* during ripple oscillations. Instead of sparse random firing, the firing of interneurons was precisely timed and locked to ripple activity. Interneurons only fired on select ripple cycles as part of an assembly of synchronized cells, thus the framework presented here may be applicable.

The mean field approach presented here, in contrast to the two stochastic one presented above, may be also applicable to populations of intrinsic pacemakers, which are abundant in the mammalian nervous system (Bean, 2024). They constitute most of the principal neurons in the basal ganglia (Surmeier et al., 2005) and cerebellum (Häusser and Clark, 1997; Raman et al., 2000), as well as cholinergic neurons in the pedunculopontine nucleus (Takakusaki and Kitai, 1997) and the striatum (Goldberg and Reynolds, 2011), suprachiasmatic neurons (Jackson et al., 2004) and several populations of slowly firing neuromodulatory neurons (Mosko and Jacobs, 1974; Williams et al., 1984; Llinas and Alonso, 1992; Ping and Shepard, 1996). Moreover, in the presence of neuromodulation, normally quiescent cells like hippocampal pyramidal neurons can be converted into pacemakers (Yamada-Hanff and Bean, 2013). In addition, sufficient background excitation can drive fluctuation-driven neurons into a fluctuation-driven regime (Schreiber et al., 2009), thus the methods described herein may have neural applications. Finally, the concepts of separating the tonic and phasic aspects of coupling interactions that vary on the same time scale as the network oscillation may apply to non-neural oscillators

**Acknowledgements**

Parallelized simulations were performed using the LSU Health HPC supercomputer *Tigerfish* obtained using NSF Grant 2018936. This work was partially funded by NIH R01NS054281 to CCC. The funding agency had no role in the collection, analysis and interpretation of data, in the writing of the report or in the decision to submit the article for publication.

**Conflict of Interest**

The authors declare no conflict of interest.

**Appendix A: Rederivation for Stability of Synchrony in an All-to-All Network with Delays**

In this section, we derive the criterion for the stability of all-to-all synchrony (Vedururu Srinivas and Canavier, 2024) based on mean field responses. In Fig. 3 of the main text, the three sets of two groups of intervals with equal length of the groups are shaded in each color (pink shade, gray shade and green box) and result in three equations:

$$ts_{N-1}[n] + tr_{N-1}[n] + ts_{N-1}[n+1] = \delta + tr_1[n] + \delta \qquad \text{(A1)}$$

$$\delta + tr_1[n] + ts_1[n+1] = ts_{N-1}[n] + tr_{N-1}[n] + \delta \qquad \text{(A2)}$$

$$\delta + tr_{N-1}[n] + \delta = ts_1[n] + tr_1[n] + ts_1[n+1] \qquad \text{(A3)}$$

These equations are the constraints that help us define the discrete map for achieving synchrony. The stability criterion is determined by the eigen values of this discrete map. The first two equations A1-A2 are sufficient to arrive at the map based on the unperturbed cluster N-1 but the third equation A3 provides the relationship between the intervals of both oscillators so that the map can be re-written using the perturbed oscillator.

Consider the equation A1. Re-writing in terms of perturbation from a steady state (denoted by an asterisk), we get

$$ts_{N-1}^* + \Delta ts_{N-1}[n] + tr_{N-1}^* + \Delta tr_{N-1}[n] + ts_{N-1}^* + \Delta ts_{N-1}[n+1] = \delta + tr_1^* + \Delta tr_1[n] + \delta$$

Removing the steady-state terms and rearranging, we get

$$\Delta ts_{N-1}[n+1] = \Delta tr_1[n] - \Delta ts_{N-1}[n] - \Delta tr_{N-1}[n]$$

Since $tr_{N-1} = P_{N-1} + M_{N-1,1}(ts_{N-1}) - ts_{N-1}$, we get $\Delta tr_{N-1} = M'_{N-1,1}(ts_{N-1}^*)\Delta ts_{N-1} - ts_{N-1}$. Using the same logic for $\Delta tr_1$and substituting for $\Delta tr_{N-1}$and $\Delta tr_1$in the above equation, we get

$$\Delta ts_{N-1}[n+1] = M_{1,N-1}(ts_1^*)\Delta ts_1[n] - \Delta ts_1[n] - \Delta ts_{N-1}[n] - M'_{N-1,1}(ts_{N-1}^*)\Delta ts_{N-1}[n] + \Delta ts_{N-1}[n]$$

Further simplifying, we get

$$\Delta ts_{N-1}[n+1] = M'_{1,N-1}(ts_1^*)\Delta ts_1[n] - \Delta ts_1[n] \ - M'_{N-1,1}(ts_{N-1}^*)\Delta ts_{N-1}[n] \qquad (A4)$$

Now, following the same procedure for the equation (A2) stepwise, we get

$$ts_1[n] + tr_1[n] + \delta = \delta + tr_{N-1}[n] + ts_{N-1}[n+1]$$

$$\Delta ts_1[n] + \Delta tr_1[n+1] = \Delta tr_{N-1}[n] + \Delta ts_{N-1}[n+1]$$

$$\Delta ts_1[n] + M'_{1,\mathrm{N}-1}(ts_1^*)\Delta ts_1[n] - \Delta ts_1[n] \ = M'_{N-1,1}(ts_{N-1}^*)\Delta ts_{N-1}[n] - \Delta ts_{N-1}[n] + \Delta ts_{N-1}[n+1]$$

$$M'_{1,N-1}(ts_1^*)\Delta ts_1[n] \ = (M'_{N-1,1}(ts_{N-1}^*) - 1)\Delta ts_{N-1}[n] + \Delta ts_{N-1}[n+1] \qquad (A5)$$

Now, substituting (A5) in (A4), we get

$$\begin{aligned}\Delta ts_{N-1}[n+1] = {} & M'_{N-1,1}(ts_{N-1}^*)\Delta ts_{N-1}[n] - \Delta ts_{N-1}[n] + \Delta ts_{N-1}[n+1] \\ & - M'_{N-1,1}(ts_{N-1}^*)\Delta ts_{N-1}/M'_{1,N-1}(ts_1^*) + \Delta ts_{N-1}[n]/M'_{1,N-1}(ts_1^*) \\ & + \Delta ts_{N-1}[n+1]/M'_{1,N-1}(ts_1^*) - M'_{N-1,1}(ts_{N-1}^*)\Delta ts_{N-1}[n]\end{aligned}$$

Rearranging further, we get

$$\begin{aligned}&\Delta ts_{N-1}[n+1]/M'_{1,N-1}(ts_1^*) \\ &\quad = -M'_{N-1,1}(ts_{N-1}^*)\Delta ts_{N-1}[n] + \Delta ts_{N-1}[n] + M'_{N-1,1}(ts_{N-1}^*)\Delta ts_{N-1}/M'_{1,N-1}(ts_1^*) \\ &\quad - \Delta ts_{N-1}[n]/M'_{1,N-1}(ts_1^*) + M'_{N-1,1}(ts_{N-1}^*)\Delta ts_{N-1}[n]\end{aligned}$$

Multiplying both sides by $M'_{1,N-1}(ts_1^*)$ and cancelling terms, we get

$$\Delta ts_{N-1}[n+1] = \left(M'_{1,\mathrm{N}-1}(ts_1^*) + M'_{N-1,1}(ts_{N-1}^*) - 1\right)\Delta ts_{N-1}[n] \qquad (A6)$$

The equation (A6) gives the discrete map in terms of intervals depending on the oscillator N-1. If we use the equation (A3) and substitute in A4 and A5, we get a similar map based on the oscillator 1. From A6, we can conclude that the system would be stable if the absolute value of the eigenvalue is less than 1, therefore, we get the criterion

$$|M'_{1,N-1}(ts_1^*) + M'_{N-1,1}(ts_{N-1}^*) - 1| < 1$$

Which can also be written as:

$$|1 - M'_{1,N-1}(ts_1^*) - M'_{N-1,1}(ts_{N-1}^*)| < 1$$

This is in the same form as the criterion derived in (Vedururu Srinivas and Canavier, 2024), but without normalization by the intrinsic period and with the response to a perturbation calculated using mean field approach rather than a single pulse.

**Appendix B: Rederivation of stability of alternating firing between two oscillators with delays.**

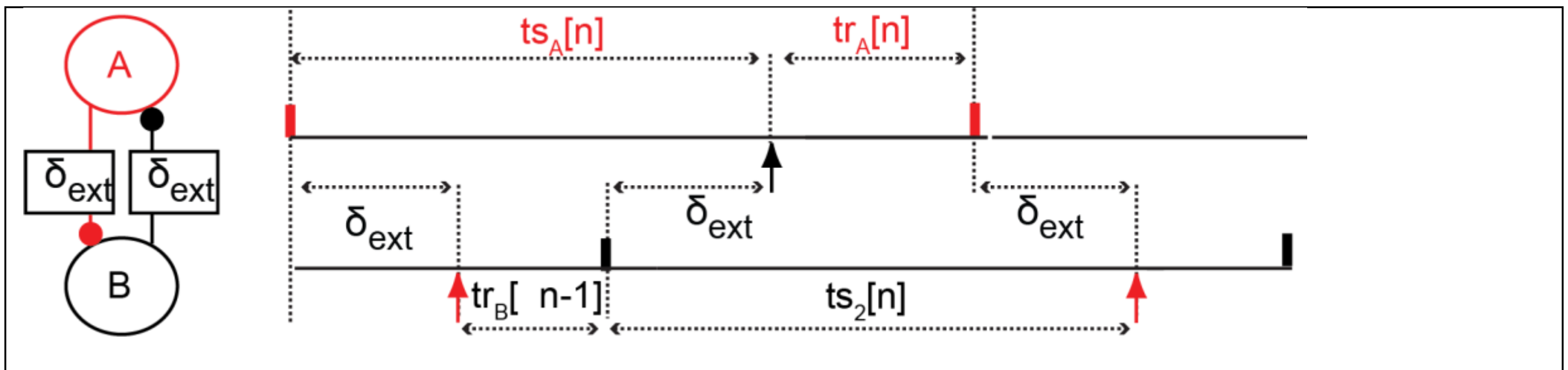

Fig B1. Discrete Map for Alternating Firing Pattern with Delays

In this section, we re-derive the criterion for determining stability of an alternating firing pattern for two oscillators, with delays, using the mean field approach. Based on the Fig. B1, the algebraic constraints of the alternating firing mode are as follows:

B1 $$ts_A[n] = 2\delta_{ext} + tr_B[n-1]$$

B2 $$ts_B[n] = 2\delta_{ext} + tr_A[n]$$

Rewriting the response intervals in terms of the mean field response

B3 $$tr_A[n] = P_{E,A} - ts_A[n] + M_{A,B}(ts_A[n])$$

B4 $$tr_B[n-1] = P_{E,B} - ts_B[n-1] + M_{B,A}(ts_B[n-1])$$

Substituting the intervals from B3 and B4 into B2 and B1 respectively, we get

B5 $$ts_A[n] = 2\delta_{ext} + P_{E,B} - ts_B[n-1] + M'_{B,A}(ts_B[n-1])$$

B6 $$ts_B[n] = 2\delta_{ext} + P_{E,A} - ts_A[n] + M'_{A,B}(ts_A[n])$$

Expand ts intervals in B5 and B6 as a perturbation from steady state

B7 $$ts_A^* + \Delta ts_A[n] = 2\delta_{ext} + P_{E,B} - ts_B^* - \Delta ts_B[n-1] + M_{B,A}(ts_B^*) + M'_{A,B}(ts_B^*)\Delta ts_B[n-1]$$

B8 $$ts_B^* + \Delta ts_B[n] = 2\delta_{ext} + P_{E,A} - ts_A^* - \Delta ts_A[n] + M_{A,B}(ts_A^*) + M'_{A,B}(ts_A^*)\Delta ts_A[n]$$

Combining B1-B3 and B2-B4: at steady state we get: $ts_x^* = 2\delta_{ext} + P_{E,X} - ts_x^* + M_{X,Y}(ts_x^*)$

We then remove steady state terms to get:

B9 $$\Delta ts_A[n] = (M'_{B,A}(ts_B^*) - 1)\Delta ts_B[n-1]$$

B10 $$\Delta ts_B[n] = (M'_{A,B}(ts_A^*) - 1)\Delta ts_A[n]$$

Substituting expression for $\Delta ts_A[n]$ from B9 into B10 we get the discrete map

B11 $$\Delta ts_B[n] = (M'_{A,B}(ts_A^*) - 1)(M'_{B,A}(ts_B^*) - 1)\Delta ts_B[n-1]$$

Based on this map, we obtain the criterion

$$\left|\left(M'_{A,B}\left(ts_A^*\right) - 1\right)\left(M'_{B,A}\left(ts_B^*\right) - 1\right)\right| < 1$$

## Appendix C: Stability of Synchrony in an Alternating Clusters

In this section, we give a brief outline for the derivation of the stability criterion in the alternating cluster case (Vedururu Srinivas and Canavier, 2024) using mean field responses approach. This is a modification of the case in Appendix B, wherein the oscillators are two cluster, but one of the oscillators is a cluster that has recurrent connections and one neuron perturbed. This results in 3 oscillators in the perturbed case. The goal is to analyze the stability of alternating firing pattern between the two clusters which also includes the synchrony within the perturbed cluster. In both the previous cases, we only had one input arriving during each cycle, which we split into the stimulus interval ts and the response interval tr. However, in the case with two clusters and a perturbed single oscillator, we have each oscillator receiving two inputs. Thus, we have two stimulus intervals and one response interval. The way to distinguish the two stimulus intervals is using the indexes $ts_{i,j}$, where i is the oscillator receiving the input and j is the oscillator eliciting an action potential. We make the assumption that the change in cycle period due to the first input occurs entirely during the second stimulus interval. Therefore, we need to correct the second stimulus intervals for each of the oscillator as follows:

$$\overline{ts}_{1,B}[n] = ts_{1,B}[n] - ts_{1,A-1}[n] + I_{1,A-1}\left(ts_{1,A-1}[n]\right)$$

$$\overline{ts}_{B,A-1}[n] = ts_{B,A-1}[n] - ts_{B,1}[n] + I_{B,1}(ts_{B,1}[n])$$

$$\overline{ts}_{A-1,B}[n] = ts_{A-1,B}[n] - ts_{A-1,1}[n] + I_{A-1,1}(ts_{A-1,1}[n])$$

Now, from Fig. 4, the sets of intervals shaded in each color (orange, pink, yellow, blue and green) have equal length. Using the corrected interval definitions above, we get following six equations:

$$2\delta + tr_{A-1}[n] = ts_{B,1}[n] + \overline{ts}_{B,A-1}[n] \qquad (C1)$$

$$ts_{B,1}[n] = \delta + tr_{A-1}[n] + ts_{A-1,1}[n+1] \qquad (C2)$$

$$\overline{ts}_{A-1,B}[n+1] = \overline{ts}_{B,A-1}[n] + tr_B[n] + \delta \qquad (C3)$$

$$\overline{ts}_{1,B}[n+1] = tr_B[n] + \delta \qquad (C4)$$

$$ts_{B,1}[n] = 2\delta + tr_1[n] \qquad (C5)$$

$$ts_{1,A-1}[n+1] = \delta + \overline{ts}_{B,A-1}[n] \qquad (C6)$$

Based on the methods in the previous sections, when we express the intervals in the above equations in terms of the mean field responses, and linearize the responses around the steady state, we get a 2D discrete system:

$$\begin{aligned}\Delta ts_{A-1,B}[n+1] \\ &= \left(M'_{A-1,B,1} - 1\right)\{M'_{B,A-1,1} + M'_{A-1,1,B} - 1\}\Delta ts_{A-1,B}[n] \\ &+ \left(M'_{1,B,A-1} - 1\right)\{M'_{B,1,A-1}(1 - M'_{B,A-1,1}) - M'_{A-1,1,B}\}\Delta ts_{1,B}[n] \qquad (C7)\end{aligned}$$

$$\begin{aligned}\Delta ts_{1,B}[n+1] = \left(M'_{A-1,B,1} - 1\right)\{M'_{B,A-1,1} - M'_{1,A-1,B}\}\Delta ts_{A-1,B}[n] + \left(M'_{1,B,A-1} - 1\right)\{\left(M'_{1,A-1,B} - 1\right) \\ - \left(M'_{B,A-1,1} - 1\right)M'_{B,1,A-1}\}\Delta ts_{1,B}[n] \qquad (C8)\end{aligned}$$

Where $M'_{i,j,k}$ is the slope of the mean field response of oscillator i due to input from oscillator j, with a background tonic input from oscillators j and k, and phasic train of input from oscillator k at the steady state. System (C7) - (C8) gives us a two-dimensional discrete linear system of the form as in (Vedururu Srinivas and Canavier, 2024). Since $M'_{1,B,A-1} = M'_{A-1,B,1} = M'_{A,B}$ as explained in the text, we can reduce these equations to the form given in the text.